\documentclass[aps,pra,twocolumn,superscriptaddress,showpacs]{revtex4}
\usepackage{graphicx}
\begin{document}
\title{Trapping of Ultracold Atoms in a Hollow Core Photonic Crystal Fiber}
\author{Caleb Christensen}
\affiliation{MIT-Harvard Center for Ultracold Atoms, Research Laboratory of
Electronics, Department of Physics, Massachusetts Institute of
Technology, Cambridge, MA, 02139, USA}
\author{Sebastian Will}
\affiliation{Institut f\"{u}r Physik, Johannes Gutenberg-Universit\"{a}t, 55099 Mainz, Germany}
\author{Michele Saba}
\affiliation{Dipartimento di Fisica, Universit\'{a} di Cagliari, Italy}
\author{Gyu-Boong Jo}
\affiliation{MIT-Harvard Center for Ultracold Atoms, Research Laboratory of
Electronics, Department of Physics, Massachusetts Institute of
Technology, Cambridge, MA, 02139, USA}
\author{Yong-Il Shin}
\affiliation{MIT-Harvard Center for Ultracold Atoms, Research Laboratory of
Electronics, Department of Physics, Massachusetts Institute of
Technology, Cambridge, MA, 02139, USA}
\author{Wolfgang Ketterle}
\affiliation{MIT-Harvard Center for Ultracold Atoms, Research Laboratory of
Electronics, Department of Physics, Massachusetts Institute of
Technology, Cambridge, MA, 02139, USA}
\author{David Pritchard}
\affiliation{MIT-Harvard Center for Ultracold Atoms, Research Laboratory of
Electronics, Department of Physics, Massachusetts Institute of
Technology, Cambridge, MA, 02139, USA}

\begin{abstract}
Ultracold sodium atoms have been trapped inside a hollow-core optical fiber.  The atoms are transferred from a free space optical dipole trap into a trap formed by a red-detuned gaussian light mode confined to the core of the fiber.  We show that at least 5\% of the atoms held initially in the free space trap can be loaded into the core of the fiber and retrieved outside.
\end{abstract}

\maketitle

\section{Introduction}
Ultracold atoms in waveguides are being used for studying quantum optics \cite{teper:cavity}, performing atom interferometry \cite{jo:recomb, wu:sagnac, wang:chip, andersson:chip}, and implementing schemes for quantum information science \cite{treutlein:chip}.  Of particular interest is the ability of waveguides and microtraps to strongly confine atoms providing high optical densities, strong interactions with light, and mechanisms for transporting atoms for further experiments.

Hollow-core optical fibers can guide and confine both atoms and light. Previous experiments have reported guiding atoms in optical dipole traps (ODTs) formed by light guided in hollow fibers \cite{renn:red, Ito:blue, dall:helium, muller:blue}; these experiments have used capillaries which guide light in multiple modes in the cladding or core.  Such fibers are susceptible to speckle or inhomogeneous fields causing uncontrolled guiding, heating or loss due to local absence of confinement.  A recently developed alternative are photonic crystal fibers which propagate a single gaussian mode confined to a hollow-core \cite{Cregan:fiber, Roberts:fiber}.  If ultracold atoms are efficiently loaded into such a mode using red detuned light, the atoms might be held for long times or controllably transported along the fiber.  The first experiments have succeeded in trapping of ultracold atoms \cite{will:thesis} or guiding of thermal \cite{takekoshi:PCF} or laser cooled atoms \cite{bajcsy:fiber} through hollow-core photonic band gap fibers.

This paper presents results so far only reported in a thesis \cite{will:thesis}.  The novelty of our results is the transfer of trapped ultra-cold sodium atoms into a hollow-core fiber, and the retrieval of a significant fraction (at least 5\%) back into the external trap.

\section{Experimental Procedure}
We produce sodium BECs using laser cooling and RF evaporation in a DC Ioffe-Pritchard magnetic trap, then load the BEC into a red detuned ODT formed by a focused laser.  The ODT focus can be moved by translating focusing optics outside the vacuum chamber~\cite{scichamb:tweezer}.  This procedure typically delivers a condensate of 10$^6$ atoms to a separate vacuum chamber holding a fiber.  The ODT is positioned 1mm from the end of a hollow-core photonic crystal fiber (2 cm long Blaze Photonics HC-1060-02) which supports a red detuned Gaussian mode in the core, hereafter called the hollow-core trap (HCT).  The fiber has a 10 $\mu$m hollow core which atoms can enter.  By adjusting the intensities of the two traps, the atoms can be controllably transferred between them.

\begin{figure}[t]
\centerline{
\includegraphics[width=8cm]{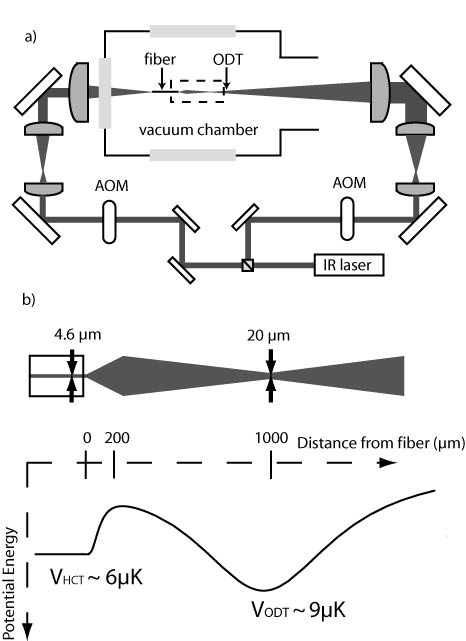}}
\label{graysetup}
\caption{(a) Diagram of the optics setup.  Note that both beams used for trapping are first order diffracted beams from acousto-optic modulators.  (b) Close up of the region near the fiber where the atoms are held and transferred.  The combined potential of the traps with 150 mW in the ODT and 5 mW in the HCT is sketched.}\label{setup}
\end{figure}

The light for the ODT and the HCT are produced by a 1064 nm Spectra-Physics J201-BL-106C diode pumped solid state multi-mode laser.  Intensities are controlled by using acousto-optical modulators. The beams are frequency shifted such that they have 50 MHz relative detuning in order to prevent static interference fringes in regions where the traps overlap.  The laser is coupled to the core mode by focusing the beam onto the fiber tip from outside the vacuum chamber (Fig. \ref{setup}).  Using a retractable mirror, the light exiting the other end of the fiber can be observed to determine how well the core mode is coupled, as well as how much light has coupled into other modes that can propagate in the cladding or on the surface of the fiber.

Because the 1064 nm trapping light is far detuned from the 589 nm $^{23}$Na $D_2$ line, the light scattering rate, which scales as $1/\delta^2$, is less than $10^{-3}$ Hz.  Therefore radiation pressure, heating and trap loss associated with scattering are negligible in the experiment.  In the far detuned limit where $\delta\gg\Omega_R,\Gamma$, the potential is given by:
\begin{equation}
U(r)=\frac{\hbar(\Omega_{R}(r))^2}{4}(\frac{1}{\omega_L-\omega_0}+\frac{1}{\omega_L+\omega_0})=\frac{\hbar\Gamma^2I(r)}{8 I_{sat}}[\frac{1}{\delta}+\frac{1}{2\omega_0+\delta}]
\end{equation}

where $(\Omega_R(r))^2$ is the squared Rabi frequency proportional to the position dependent beam intensity I(r), $\delta=\omega_L-\omega_0$ is the laser detuning in radians/sec, and $\Gamma$ is the natural linewidth.  For the sodium \textit{D$_2$} line, saturation intensity I$_{sat}$=6 mW/cm$^2$ and $\Gamma=2\pi\times10$ MHz.  Note that the counter-rotating term accounts for 25\% of the potential.

The focus of the ODT has a waist \textit{w$_0$} = 20 $\mu$m, approximated by a gaussian profile of
\begin{equation}
I(\rho,z)=\frac{2P}{\pi w(z)^2}e^{-2r^2/w(z)^2}
\end{equation}
where $\rho$ is the radial coordinate, z is axial distance from the focal plane, \textit{w(z)} is the beam radius $w(z)=w_0\sqrt{1+(\frac{\lambda z}{\pi w_0^2})^2}$, and \textit{P} is the beam power.  The HCT mode is approximately gaussian in the radial direction with waist of 4.6 $\mu$m.  Assuming only the core mode is excited, the mode is axially constant along the fiber and diverges at the fiber tip according to the gaussian formula, with the appropriate \textit{P} and \textit{w$_0$}.

The calculated depths of the ODT and HCT are 5.8 x 10$^{-2}$ $\mu$K/mW and 1.2 $\mu$K/mW respectively (Fig. \ref{setup}b).  The maximum power in the ODT is 150 mW.  Potentials are qualitatively consistent with our observations by assuming a maximum power of 5 mW in the core mode of the fiber.  Bench tests suggested three times more power could couple through the fiber, but we believe that the extra power was in surface or cladding modes and didn't contribute to the core intensity.

\begin{figure}[t]
\centerline{
\includegraphics[width=8cm]{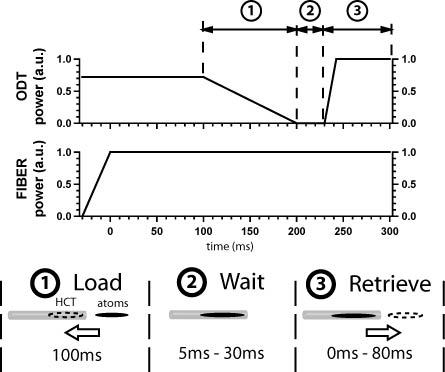}}
\caption{Timeline for transferring atoms by ramping laser power.  The fiber light is ramped up while the atoms are in the ODT, and the ODT is ramped down to transfer to the HCT (1), turned off for holding (2) and ramped up to retrieve the atoms(3).}\label{ramps}
\end{figure}

The ODT is positioned in front of the fiber, and the laser intensities in the two traps are ramped to perform the transfer (Fig. \ref{ramps}).
As the ODT depth is reduced (step 1 in the figure), atoms are pulled out of the trap and are accelerated into the potential well of the HCT.  The ODT is turned off completely while the atoms are held within the HCT (step 2).  After some hold time the ODT is ramped up and atoms transfer back to the ODT (step 3).

Absorption images are obtained during the loading and retrieving process to measure the atom number in the ODT (Fig. \ref{images}).  From the final ODT number we determine the overall efficiency of the process.

\section{Results}
Based on the analysis of absorption images we measure 5 x 10$^4$ atoms in the ODT after retrieval, corresponding to about 5\% of the 10$^6$ atoms initially delivered at the start of the experiment.  The successful transfer of atoms into the HCT and their retrieval is the main result of this paper.

\begin{figure}[t]
\centerline{
\includegraphics[width=8cm]{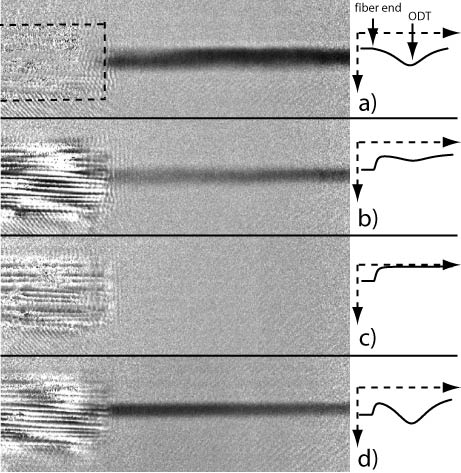}}
\caption{Images of atoms in the ODT during the experiment.  Also shown are sketches of the corresponding combined potential of the HCT and ODT.  (a) Atoms are held in the ODT near the fiber, with no light coupled into the fiber.  The dashed line indicates the position of the 100 micron thick fiber.  (b) Light is coupled to the fiber, and as the ODT intensity is ramped down, atoms are depleted from the ODT until (c) no atoms remain outside the fiber when the ODT power reaches zero.  (d) After ramping the ODT back up, atoms that were trapped in the HCT return to the ODT.}
\label{images}
\end{figure}

Images recorded while the ODT power is ramped down reveal that the number of atoms gradually reduces until the ODT power reaches zero (see Figs. \ref{images}b and \ref{images}c).  Once the ODT power drops to zero, any remaining atoms either fall into the HCT potential well, or are lost from either trap.

The atoms are then held in the trap for up to 30 ms, several hundred radial trapping periods. After this time, the ODT intensity is ramped back up, and the atom number outside the fiber increases over 80 ms (Fig. \ref{images}d) until a maximum is reached.  In varying the hold time, the retrieved number changed by at most 30\%, indicating lifetime longer than 50 ms.  Longer hold times were not explored in the current setup.

To rule out alternate trapping mechanisms, we performed some experimental runs under identical parameters except with no light in the HCT, and found that no atoms are recovered, indicating that the atoms indeed are loaded into and retrieved from the HCT.

The initial and final cloud sizes were similar, implying similar temperatures before and after the transfer.  This temperature may be estimated at a few $\mu$K by assuming evaporation continually occurs in the finite depth optical trap, leading to temperatures at a fraction of the trap depth.

\section{Discussion}

Our experimental scheme for transferring atoms between the two traps was guided by the concept of adiabatic transfer in a double well potential, where the atoms always occupy the deeper well.  Our results are consistent with this picture.  We do not know whether the transfer of atoms involves tunneling, spilling over the barrier, or sloshing. This reflects that the potential between the two traps is likely to be affected by spurious modes traveling along the fiber surface or in the cladding and interfering with the light in the core.  In fact, we frequently observed light coupled into the cladding leading to distortions in the mode after the fiber.  Without proper alignment, the cladding modes prevented loading atoms into the fiber or trapped atoms in local maxima outside the fiber.  The successful transfer was possible only with great care in coupling to minimize light in the cladding modes.  In the future, a coupling lens inside the vacuum could replace the final lens (which was 10 cm away from the fiber) resulting in more controlled and stable coupling of light.  We do not try to describe the exact dynamics of the transfer without better knowledge of the potential between the two wells.

Observation of the atoms inside the fiber would reveal the dynamics of the loading and trapping process.  Although the fiber is transparent to resonant light, absorption imaging from the side was not possible due to severe scattering and refracting.  However, resonant light could be propagated along the core mode to provide information on the integrated density of atoms inside the fiber.

We were unable to observe atoms being guided completely through the fiber, probably because of low atom density and the lack of an appropriate trap in which to collect them after the fiber.  It may turn out to be difficult to control the light intensity along the fiber core sufficiently to avoid undesirable accelerations of the atoms, to localize them inside the fiber or to propagate them through the fiber in a controlled way.  In this case, it may be advantageous to use the fiber mode for strong transverse confinement but add magnetic confinement for the axial direction.  This could be accomplished with a  quadrupole trap, where by changing the balance of currents in anti-Helmholtz coils, the axial minimum could be swept across the fiber to controllably propagate the atomic cloud.  With this setup, one could obtain detailed information on atom lifetime at different positions within the fiber.  Originally, we had the intention to take more quantitative data with such an improved setup, but the priorities of the lab changed.  Therefore, we have presented our qualitative results in this paper.

In conclusion, we have shown that optical dipole traps are well suited to load ultracold atoms into a hollow-core photonic crystal fiber.  The reported retrieval efficiency of 5\% is a lower bound for the transfer efficiency, and can probably be substantially increased with an improved setup.

This work was funded by DARPA, NSF, and ONR.  G. -B. Jo and S. Will acknowledge additional support from Samsung Foundation and the Studienstiftung des Deutschen Volkes, respectively.

\end{document}